# Fingerprint of QPT in a bilayer-quantum-well system at the filling fraction ν = 5/2 in the low temperature range (1K-100 K)

Partha Goswami

Physics Department, D.B.College( University of Delhi), Kalkaji, New Delhi-110019, India
*Email id of communicating author: physicsgoswami@gmail.com

**Abstract** We consider the spin polarized fermions for the filling fraction 5/2 in a bi-layer quantum well system. Since the kinetic energy of the system in fractional quantum Hall states (FQHS) is totally quenched, the Hamiltonian describing the system comprises of the electron correlation and tunneling terms. The correlations are captured by the 'so-called' Haldane pseudo-potentials(HPP). We employ the finite-temperature formalism involving Matsubara propagators to deal with this Hamiltonian. We show that the system undergoes a zero-order quantum phase transition (QPT), at fixed charge imbalance regulatory(CIR) parameter and constant layer separation as the inter-layer tunneling (ILT)strength is increased, from the effective two-component state (two independent layers) to an effective single-component state (practically a single layer). At finite and constant ILT strength, a transition from the latter state to the former state is also possible upon increasing CIR parameter. We identify the order parameter to describe this QPT as a pseudo-spin component and calculate the order parameter with the aid of the Matsubara propagators. The clear finger-print of this QPT is obtained up to temperature equal to 100 K.

We visualize a bi-layer system as the one consisting of two parallel (quasi-)two-dimensional electron systems of width 'w' separated from one another by a tunnelling barrier of thickness d (d ≥ (w/2)). The barrier height and thickness can be adjusted such that electrons are either localized in separate layers or delocalized between the two layers. We consider spin-less fermions for the filling fraction ν = 5/2 confined in this planar geometry. The well and the barrier materials, respectively, are assumed to be GaAs and AlGaAs. The Hamiltonian, in the symmetric-anti-symmetric basis (SAS) and expressed in units of ($e^2/\varepsilon l_B$) where the magnetic length $l_B = \sqrt{(\hbar/eB)}$ is the length unit and ε is the permittivity of the system material, comprises of the electron correlation (captured by the 'so-called' Haldane pseudo-potentials), the inter-layer tunneling (ILT), and the charge imbalance regulatory(CIR) terms. The reason for considering the last two terms is that the recent experiments [1,2] have achieved bi-layer fractional quantum Hall(FQH) systems where both the inter-layer and charge imbalance tunneling terms can be controlled by changing system parameters such as gate voltages. The kinetic energy of the system in the fractional quantum Hall states (FQHS) is totally quenched. Under the independent Landau level(LL) assumption, we first show that, at the temperature ~ 50 mK, the system undergoes a zero-order quantum phase transition (QPT) at fixed CIR parameter Δ and the constant layer separation, as the ILT strength $\Delta_{SAS}$ is increased, from the effective two-component state (bi-layer fractional quantum Hall state(FQHS)) to an effective single-component state (single layer FQHS); at finite and constant $\Delta_{SAS}$ a transition from the latter state to the former state is also possible upon increasing Δ (see Figure 1). We identify the order parameter (Γ) to describe the QPT to be a pseudo-spin component involving the inter-layer tunneling and analogous to the component $S^z$ of the single spin-1/2 operator **S**.

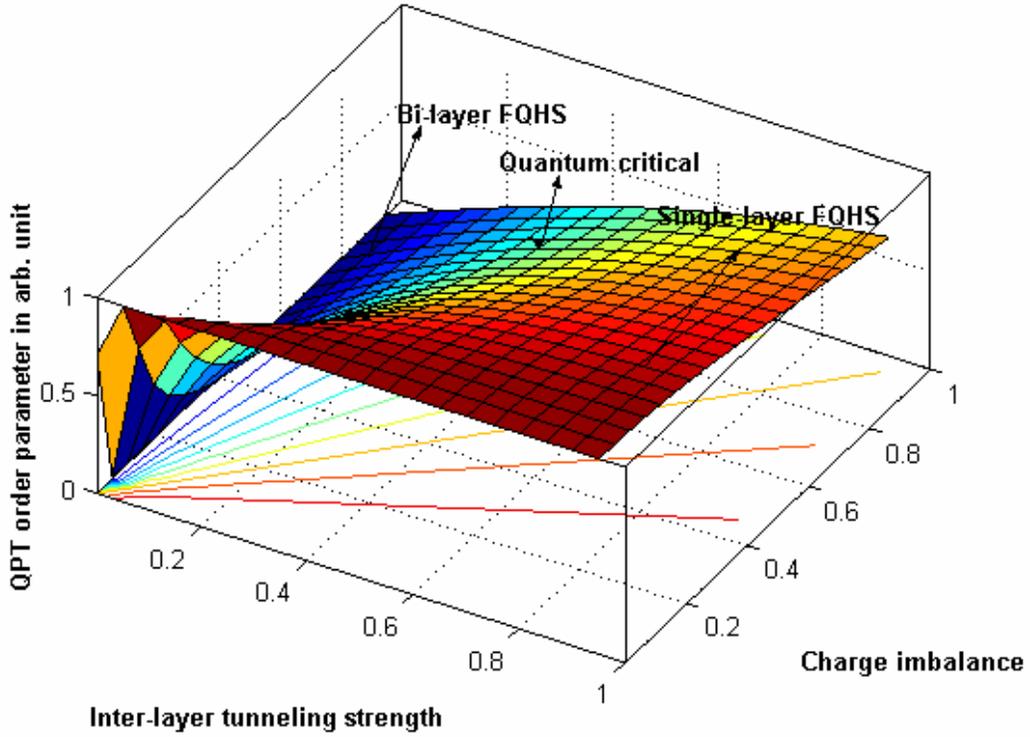

**Figure 1.** A 3-D plot of the QPT order parameter on the $\Delta_{SAS}$ - $\Delta$ zone at the temperature T = 50 mK for the spin polarized fermions at the filling fraction $\nu$ = 5/2. We show that, in the $\nu$ = 5/2 case, a bi-layer system undergoes a zero-order quantum phase transition (QPT), at fixed charge imbalance regulatory parameter $\Delta$ and constant layer separation as the inter-layer tunneling strength $\Delta_{SAS}$ is increased, from the effective two-component state (two independent layers) to an effective single-component state (practically a single layer). At finite and constant $\Delta_{SAS}$ a transition from the latter state to the former state is also possible upon increasing $\Delta$. In the previous theoretical works[3,4,5,6], the former state has been linked with the Halperin Abelian 331 FQHS and the latter with the non-Abelian Moore-Read Pfaffian FQHS. It is also reported that theoretically it is not possible for the Pf FQHS to exist in the LLL, and therefore this type of phase transition is unlikely for the $\nu$ = ½ case.

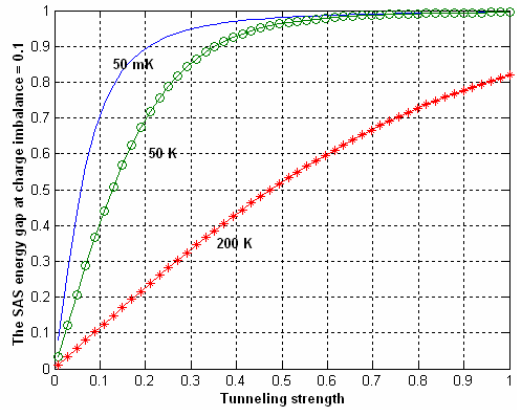

**Figure 2.** A 2-D plot of the SAS energy gap (or the QPT order parameter $\Gamma$ ) as a function of $\Delta_{SAS}$ for $\Delta$ = 0.1 at the temperatures T = 50 mK, 50 K and 200 K for the spin polarized fermions at the filling fraction $\nu$ = 5/2 in the weak correlation case $(d/l_B)$ = 1.5. The softening of all of the pseudo-potentials occurs for $(d/l_B)$ > 1. We find that there is a discontinuity in the susceptibility $\chi \equiv (\delta\Gamma/ \delta\Delta_{SAS})$ while the system undergoes

transition from the bi-layer FQHS to the single-layer FQHS. Whereas the plot at 50 K carries the finger-print of QPT mentioned above, at T = 200 K the finger-print is completely obliterated. The plot indicates that the quantum Hall effect related phenomena are strongly affected by the increase in temperature.

It must be noted that we did not opt for the zero-temperature wave-function approach to arrive at this QPT. We have employed the finite-temperature formalism involving Matsubara propagators to deal with the Hamiltonian and calculate the order parameter with the aid of these propagators. We have been able to obtain the finger-print of this QPT at finite but low temperature (see the curves for 50 mK and 50 K in Figure 2) and interpret the outcomes in the language of the thermal phase-transition (TPT). For example, as the entropy is of increasing importance in TPT for determining the phase of systems with rising temperatures T, for the QPT here the term $\Gamma$ is to be accorded a similar status vis-à-vis increasing $\Delta_{SAS}$. We note that $\Gamma$ corresponds to an analytically tractable quantity - the symmetric-anti-symmetric(SAS) single-particle excitation spectrum gap which is found to be an increasing function of $\Delta_{SAS}$ for the given Coulomb repulsions (see Figure 2). The counterpart of the specific heat capacity here is the susceptibility $\chi \equiv (\delta\Gamma/ \delta\Delta_{SAS})$. Analogous to the second-order thermal phase transition, we conclude from Fig.2 that there is a discontinuity in $\chi$ while the system undergoes transition from the bi-layer FQHS to the single-layer FQHS.

In conclusion, the semi-conductor bi-layers with finite single-layer width can support both the Halperin Abelian 331 and the non-Abelian Moore-Read Pfaffian like FQHS for the second Landau level and that there could be a QPT, as a function of tunneling strengths (at constant layer separation), between these states. The discernable signature of this QPT could be obtained up to temperature T ~ 100 K as shown in Fig.2. This is an experimentally verifiable prediction of our finite-temperature approach.